\documentclass[manuscript,useAMS,usenatbib]{aastex}

\shorttitle{Detections of X-ray emissions from Type Ia Supernova 2003lx}
\shortauthors{Kwan Lok, Li}

\begin{document}
\title{Detections of X-ray emissions from Type Ia Supernova 2003lx}
\author{K.L. Li, Chun.S.J. Pun}
\affil{Department of Physics, University of Hong Kong, Pokfulam Road, Hong Kong, PR China}

\begin{abstract}
We present a study of a young (few years old) supernova remnant 2003lx which was first discovered in X-ray through two serendipitous \textit{Swift\/} observations in 2008 January and the corresponding merged image revealed a 7$\sigma$ source detection at $2.1\pm1.2$ arcsec ($0.9 \pm 0.5$~\textit{Swift\/} pixels) from the optical position of the supernova. The X-ray luminosity $L_x=4.8_{-1.7}^{+1.8} \times 10^{41}$~ergs s$^{-1}$ of band 0.3 - 2.0 keV is estimated by a redshift $z = 0.0377$ power law which can infer a companion star with a mass-loss rate of $\dot{M} \sim 10^{-4} M_\sun$ yr$^{-1}$ (with assumption of wind velocity $v_w=10$ km s$^{-1}$) in the white dwarf binary system. Thermal Model fitting suggests the temperature of the shock wave front is kT $\sim 0.4$ keV which is consistent with the typical reverse shock temperate. The X-ray emission allows us to probe the interaction between the fast moving debris of the exploded star and the circumstellar medium (CSM). This in turns provides footprints of the mass-loss history of the exploded system, and may allow us to learn about the nature of the progenitor and its companion. \\
\end{abstract}


\section{Introduction}
Type Ia supernovae (SNe Ia) historically have been classified by the absence of hydrogen lines in their spectra. 
They are generally believed to be arising from explosions of white dwarf binary systems when the white dwarf accretes mass from the companion up to the Chandrasekhar limit. However, the exact nature of the progenitor remains largely unknown, particularly on the nature of the companion, with single-degenerate (companion a main-sequence star or red giant) or double-degenerate scenarios (companion another white dwarf) still under debate. In case for single-degenerate models, it is possible that some hydrogen-rich materials remain in the circumstellar materials (CSM) after the supernova explosion. This hydrogen was first observed in SN~2002ic, a normal SN~Ia at the time of discovery, but later developed a broad H$\alpha$ (FWHM $\sim$ 1800 $\mathrm{km\,s^{-1}}$) component in the spectrum, believed to be arising from the interaction of the supernova debris with the dense CSM \citep{2002ic}. This could suggest that at least some Type~Ia have stellar companion (though some have argued that the SN is a actually Type~Ic that has gone through heavy mass loss in a gas-rich environment, see \citet{2002ic_Ic}). A similar example, SN~2005gj, was found to closely resemble SN~2002ic \citep{2005gj}.

The shock interaction between the fast expanding supernova envelope and the CSM leads to vast amount of X-ray, optical, and radio emissions. Thermal X-ray emission is expected to arise from the hot gas in the interaction region between the ejecta and the materials from the mass loss before the supernova explosion. X-ray emission from supernovae thus offers a channel to probe the properties and structure of the CSM. The list of X-ray supernovae observed has grown extensively with missions such as \textit{Swift\/} since its launch in 2004. As expected, a majority of the 47 candidates listed in the online \textit{Swift\/} X-ray supernova catalog\footnote{http://lheawww.gsfc.nasa.gov/users/immler/supernovae\_list.html} maintained by S.~Immler are core collapse supernovae under different categories (SN~II, IIP, IIn, IIb, Ibc). The only one exception is the Type Ia SN~2005ke in NGC 1371, of which a $\sim$~3~$\sigma$ detection had been reported by Immler et al. (2006) from a merged 267.8~ks observation taken by \textit{Swift\/} at $5-132$~days after explosion. The reported $0.3-2$~keV unabsorbed flux is $\sim 4 \times 10^{-15}$~ergs cm$^{-2}$ s$^{-1}$. If the result is correct, this would have been the first detected X-ray Type Ia superonva. On the other hand, the same set of \textit{Swift\/} data (together with some late time \textit{Chandra\/} ACIS data taken 110 days after explosion) was reanalyzed by \citep{chan_typeIsn}) and they concluded that no photons were detected within 14$"$ of the SN~2005ke position. Moreover, neither SN~2002ic nor SN~2005gj had been detected in X-ray. 

We has recently completed a study of the entire \textit{Swift\/} archive to search for X-ray supernovae (which have been discussed in Paper I \footnote{\textit{Swift} detection of X-ray supernovae which will be submitted to MNRAS}). We adopted an automated search of the \textit{Swift\/} data up to September 2010 and identified a total of 23 X-ray SNe candidates. While most of our candidates are identical to those listed in the online catalog, we were not able to detect 5 of the previously published candidates, including SN~2005ke. We believe this is probably due to our searching strategy of doing a blind search instead of a targeted search. We therefore believe the detection of the first Type Ia X-ray SN to be uncertain at this stage. 

\begin{figure}
\center
\includegraphics[width=70mm]{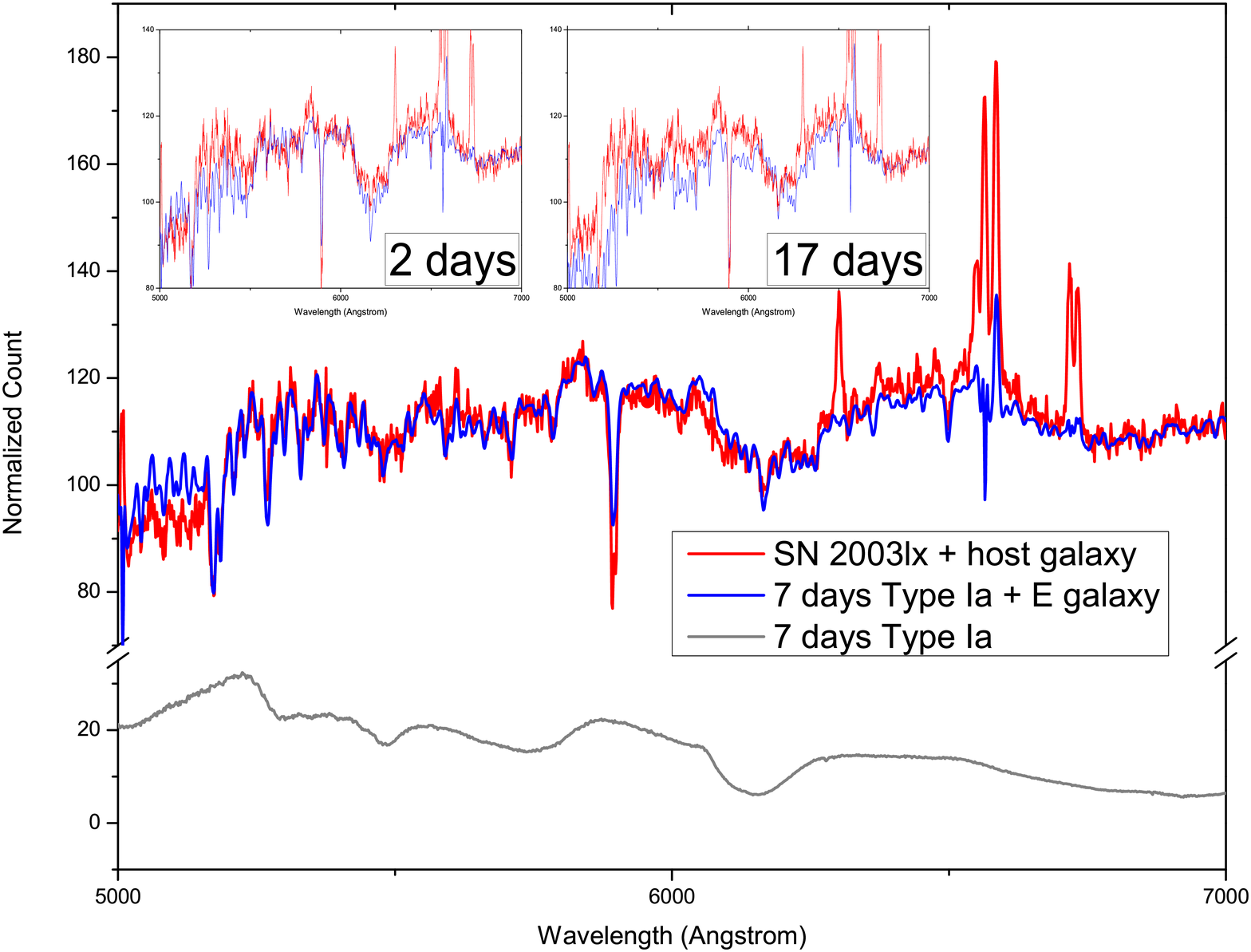}
\caption{\footnotesize{The SDSS discovery spectrum of SN~2003lx (red line) is plotted along with a 7-days after maximum spectrum of Type Ia SN~1994D (grey) obtained from the SUSPECT archive. The combined elliptical galaxy template spectrum + SN spectrum (blue) is shown to fit the SDSS data well.}}
\label{sdss-spectrum}
\end{figure}

\section{Observations and Data Analysis}
SN~2003lx is a Type Ia discovered by the Sloan Digital Sky Survey (SDSS) on 2003 April 24 \citep{2003lx} from an anonymous host galaxy at a distance 155~Mpc (redshift z = 0.0377, $H_0$ = 71 km~s$^{-1}$ Mpc$^{-1}$). The supernova was identified spectroscopically by subtracting the underlying galaxy spectrum using a method based on Principal Component Analysis \citep{SDSS_Ia_SNe}. The age of the supernova was estimated to be 10 days after maximum by fitting supernova spectral templates. The discovery SDSS spectrum of SN~2003lx is shown in Figure~\ref{sdss-spectrum}, together with an elliptical galaxy template and a 7 days after maximum template of SN~Ia from the SUSPECT spectrum archive\footnote{http://suspect.nhn.ou.edu/}. It can be seen that the observed SN~2003lx spectrum can be well-fitted, in particular the absorption feature near 6150~\AA\ due to Si~II~$\lambda6355$. We are therefore confident that SN~2003lx is indeed a Type~Ia supernova, even though no photometry of the supernova is available.

\subsection{UV/Optical emssion}
The \textit{Swift UVOT\/} observed the counterpart of the X-ray source in UV/optical band. We used the standard task \textit{uvotsource} with a 5 \arcsec radius circle extraction region to extract the optical/UV magnitudes. The UV/optical magnitudes of it are v = 15.71, b = 16.78, u = 17.31, uvw1 = 18.49, uvm2 = 19.64 and uvw2 = 19.42. Using the color transformation equations done by \citet{uvot}, the magnitudes in Johnson filters system are V = 15.71, B = 16.78 and U = 17.16. In order to find the cross identifications of the source. A five arcsec radius search was carry out in \textit{2MASS} and USNO-B1.0 catalogs and corresponding IR source 2MASS-J16192164+4105237 and USNO-B1.0-1310-0266055 were found in the position of SN 2003lx with a offset of $0.04 \arcsec$. 

\begin{figure*}
\center
\includegraphics[width=150mm]{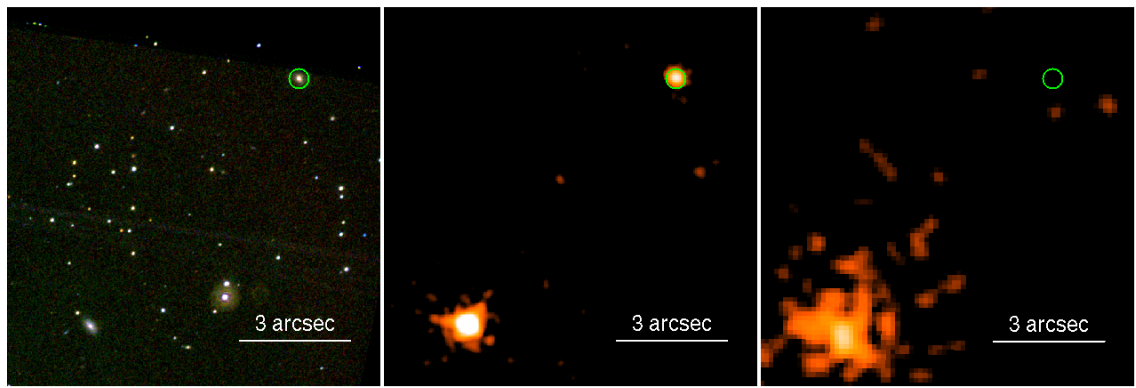}
\caption{\footnotesize
{Aligned Soft band (0.3-2.4 keV) X-ray image of SN~2003lx, taken by \textit{Swift UVOT\/} (Left), \textit{Swift XRT\/} (middle) and \textit{ROSAT\/} (Right)}}
\label{2003lx-swift}
\end{figure*}

\begin{figure}
\center
\includegraphics[width=70mm]{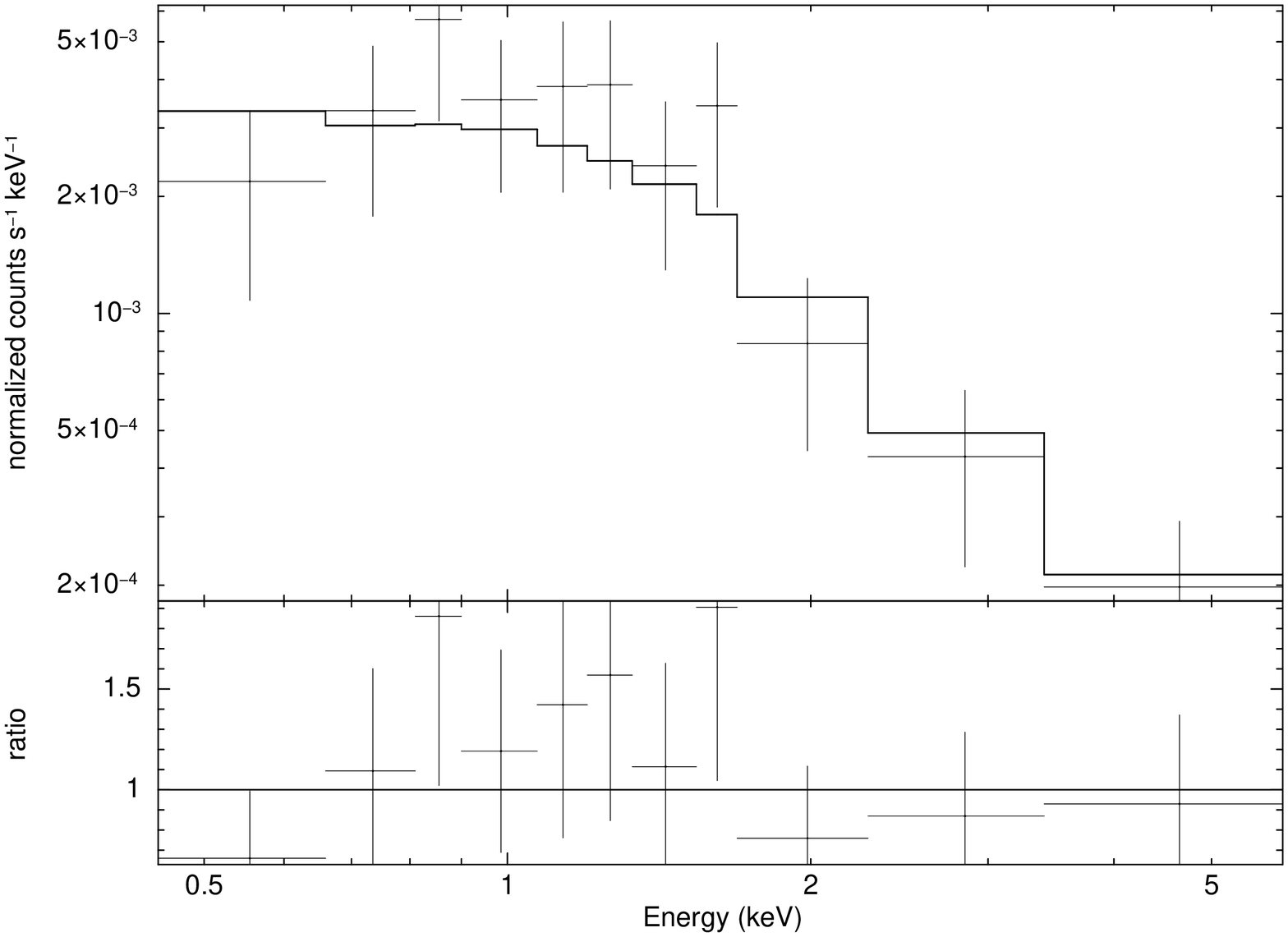}
\caption{\footnotesize
{X-ray spectrum with Power Law model of index 1.8}}
\label{powlaw}
\end{figure}

\subsection{X-ray emssion}
The field of SN~2003lx was serendipitously imaged by the XRT on board of \textit{Swift\/} twice in 2008 January, with a total exposure time of 14~ks.  The observation log is presented in Table~\ref{tab-obs}.  As shown in Figure~\ref{2003lx-swift}, the position of the supernova is near the edge of the XRT field-of-view. A source is visible in both observations with the count rate determined from merging two data sets to be $(5.7\pm0.8) \times10^{-3}\,\mathrm{count\,s^{-1}}$, corresponding to a $7 \sigma$ detection. With the center of the X-ray source at only $2.1 \pm 1.2$~arcsec ($0.9 \pm 0.5$~pixels, using \textit{Swift\/} coordinates) away from the optical position of the supernova, we believe the discovered X-ray source is indeed either SN~2003lx or a system that is closely associated with the supernova. Assuming a power-law model with a Galactic hydrogen column density\footnote{nH value from http://heasarc.nasa.gov/cgi-bin/Tools/w3nh/w3nh.pl} $N_{\rm H} = 8.44\times10^{19}\,\mathrm{cm^{-2}}$, the best spectral fitting suggests $\Gamma = 1.8_{-0.3}^{+0.3}$ and  we derive the unabsorbed $0.3-2.0$ keV flux of the source to be ${1.7_{-0.6}^{+0.6}\times10^{-13}}$ erg cm${^{-2}}$\,s${^{-1}}$, corresponding to a X-ray luminosity of $L_{\rm 0.3-2.0 keV} = 4.8_{-1.7}^{+1.8} \times 10^{41}$~ergs s${^{-1}}$ assuming a distance to the source galaxy at 155~Mpc. We also fitted the spectrum with a thermal model and the best fit value indicated that the temperature of the object is $\sim 0.4$ keV. 

\begin{table}\footnotesize
\label{tab-obs}
\begin{minipage}{70mm}
\caption{\textit{Swift} Observations of SN 2003lx}
\begin{tabular}{@{}cccc}
\hline
OBS-ID & DATE & EXP & COUNT RATE \\ 
& (UT) & (s) & ($10^{-3}\,\mathrm{count\,s^{-1}}$) \\ 
\hline
00036548001 & 2008-01-10 20:39:16 & 2330 & $4.2\pm2.1$\\
00036548002 & 2008-01-14 11:16:45 & 11480 & $5.9\pm0.9$\\
\hline
\end{tabular}
\end{minipage}
\end{table}

We also investigated archival image of the field taken by the \textit{ROSAT\/} satellite (Figure~\ref{2003lx-swift}). The same field had been monitored by the PSPC scan mode on 1990 July 30 for 760~seconds. Assuming the same value of $\Gamma$ and $N_{\rm H}$ as above, we calcluated at the newly-discovered \textit{Swift} source should yield a total of 13 counts in \textit{ROSAT}. However, only 3 net counts were detected (total 38 counts and 35 background counts) in the 1990 image. While the high background does reduce the significance of the result, we are still able to exclude the presence of the source at 91.2\% confidence level in the \textit{ROSAT} data. In addition, An upper limit estimation method introduced by \citet{chan_typeIsn} was used to calculate the upper limit flux, the upper limit from the position in 1990 is about ${\sim 10^{-14}}$ erg cm${^{-2}}$\,s${^{-1}}$ which is much dimmer than the \textit{Swift\/} detection in 2008. The positional coincidence and the absence of an earlier X-ray source strongly suggest that the discovered \textit{Swift} source is indeed a X-ray Type~Ia supernova. 



\section{Discussion}
The XRS located at a anonymous galaxy of ~155 Mpc distance from us. The high X-ray flux number (${\sim 10^{-13}}$ erg cm${^{-2}}$\,s${^{-1}}$) detected by \textit{Swift\/} suggests the corresponding X-ray luminosity of ($\sim 10^{41}$~ergs s${^{-1}}$) is relatively high to typical upper limit estimations of type Ia \citep{chan_typeIsn}. The dim X-ray upper limit flux (${\sim 10^{-14}}$ erg cm${^{-2}}$\,s${^{-1}}$) detected by \textit{ROSAT\/} shows that there was no X-ray source at the emission position in 1990. However, it is still possible to say the XRS could be a high variable AGN. In order to justify the the identification of it, we used the criteria developed by \citet{AGN_cri} to check. The criteria including (a) \textit{X-ray Luminosity:} Sources with ${L_x \geq 3 \times 10^{42}}$~ergs s${^{-1}}$ usually do not belong to purely star-forming non-AGN galaxyies which mean for those source that the luminosities are greater than this value could be classified as a luminous AGN. (b) \textit{X-ray Hardness:} A high absorption ($nH \gtrsim 10^{22}\,cm^{-2}$) would lowers a AGN source observed flux to a level below the critical value. Meanwhile, the observed spectrum of it should be extremely high (effective power index of $\Gamma_\mathrm{eff} \leq 1.0$), this kind of sources can be flagged as obscured AGNs. (c) \textit{X-ray-to-SFR Correlation:} Using  $\mathrm{SFR}(M_\sun\,\mathrm{yr}^{-1}) = 9.8 \times 10^{-11} (L_\mathrm{UV}/L_\sun + L_\mathrm{IR}/L_\sun)$ to estimate the star-formation rate (SFR). If a source has $L_\mathrm{0.5-8keV}/\mathrm{SFR} \gtrsim 3 \times (1.06\times10^{40}$~ergs s${^{-1}}\,{(M_\sun\, \mathrm{yr}^{-1})^{-1}})$, we classify it as an AGN. (d) \textit{X-ray-to-Optical Flux Ratio:} The X-ray-to-optical flux ratio log($f_\mathrm{0.5 keV,int} / f_\mathrm{R} ) = -1$ is a useful discriminator between AGN and galaxy. A source with log($f_\mathrm{0.5 keV,int} / f_\mathrm{R} ) > -1$ is mark as an AGN. 

For the XRS, the estimated flux is about $\sim 5\times10^{41}$~ergs s$^{-1}$, which is obviously greater than the critical value set by Criterion (a). The effective power law index of the source is about 2.0 which is significant soft to reject the possibility that it is be an obscured AGN. Although we do not have an IR ($\sim \mu$m in wavelength) counterpart to see if the XRS passes the Criterion (c), we have a r-band magnitude 14.13 from SDSS\footnote{http://cas.sdss.org/astrodr7/en/tools/explore/obj.asp?id=\\587733609089400991} to test the Criterion (d). We first transform the SDSS r-band magnitude into Johnson R-band filter \footnote{http://www.sdss.org/dr7/algorithms/sdssUBVRITransform.html} which gives R-band magnitude ~13.99 corresponding to a $f_\mathrm{R}$ of $8.5\times10^{-12}$ erg cm${^{-2}}$\,s${^{-1}}$. The calculated value of X-ray optical flux ratio is $-1.45 < -1$. Three out of four criteria suggest the XRS is not likely to be a typical AGN, so we concluded that the host galaxy is not the main source of the strong X-ray flux. 

A more sound explanation of the X-ray emission may be the circumstellar interaction that came from the mass lost by the companion of SN 2003lx. In order to understand the CSM environment that can produced such a high luminosity, we use model of hydrodynamic interaction for a typical Ia to estimate the mass loss rate of the companion \citep{2005ke}. Assuming a constant mass-loss rate $\dot{M}$ and wind speed $v_w$ .The luminosity in the forward shock region is $L_x=1/(\pi m^2)\Lambda (T)(\dot{M}/v_w)^2(v_st)^{-1}$, where $m$ is the mean mass per particle (assume a H+He plasma, $2.1\times10^{-24}$) and $\Lambda (T)$ is the cooling function of the plasma with respect to temperature $T$. Putting $\Lambda_\mathrm{0.3-2.0 keV} = 3 \times 10^{-23} \mathrm{erg\,cm^3\,s^{-1}}$, $L_\mathrm{reverse} = 100\,L_\mathrm{forward}$ and a forward shock velocity $v_s = 10000\,km\,s^{-1}$ \citep{CF82,inter_1993J}, a mass-loss rate of $\dot{M} \sim 3\times10^{-4}\,M_\sun\,\mathrm{yr}^{-1}(v_w/10\,\mathrm{km\,s^{-1}})$ is calculated. This high mass-loss rate with low wind speed suggests a dense CSM environment around the SN. Shock wave interacts with such high dense CSM produces the high observable X-ray flux. Similar CSM of mass-loss rate of $\dot{M} \sim \times10^{-3}\,M_\sun\,\mathrm{yr}^{-1}(v_w/10\,\mathrm{km\,s^{-1}})$ were also found in SN 2002ic by \citet{1.5SN} where the CSM may be a product formed by a slow and dense stellar wind from a super giant companion star like VY CMa \citep{red_mass_loss}.

This work is based on observations obtained with \textit{Swift}, a part of NASA's medium explorer program, with the hardware being developed by an international team from the United States, the United Kingdom and Italy, with additional scientific involvement by France, Japan, Germany, Denmark, Spain, and South Africa. K.L. Li acknowledges support from the University of Hong Kong under the giant of Postgraduate Studentship. C.S.J. Pun acknowledges support of a RGC grant from the government of Hong Kong SAR.

\end{document}